\documentclass[12pt]{article}
\usepackage{epsfig}
 
\newcommand{\mrm}[1]{\mathrm{#1}}
\renewcommand{\c}{\mrm{c}}
\renewcommand{\d}{\mrm{d}}
\newcommand{\e}{\mrm{e}}
\newcommand{\g}{\mrm{g}}
\newcommand{\p}{\mrm{p}}
\newcommand{\q}{\mrm{q}}

\renewcommand{\u}{\mrm{u}}
\newcommand{\cbar}{\overline{\mrm{c}}}
\newcommand{\dbar}{\overline{\mrm{d}}}

\newcommand{\qbar}{\overline{\mrm{q}}}

\newcommand{\ubar}{\overline{\mrm{u}}}
 
\newcommand{\lessim}{\raisebox{-0.8mm}%
{\hspace{1mm}$\stackrel{<}{\sim}$\hspace{1mm}}}
\newcommand{\alphas}{\alpha_{\mrm{s}}}
\newcommand{\alphaem}{\alpha_{\mrm{em}}}
 
\newenvironment{Itemize}{\begin{list}{$\bullet$}%
{\setlength{\topsep}{0.2mm}\setlength{\partopsep}{0.2mm}%
\setlength{\itemsep}{0.2mm}\setlength{\parsep}{0.2mm}}}%
{\end{list}}
\newcounter{enumct}

\newlength{\abstwidth}
\begin{document}
 
\sloppy
 
\pagestyle{empty}
 
\begin{flushright}
CERN--TH/96--04  \\
LU TP 96--2 \\
January 1996
\end{flushright}
 
\vspace{\fill}
 
\begin{center}
{\LARGE\bf Parton Distributions of the Virtual Photon}\\[10mm]
{\Large Gerhard A. Schuler$^a$} \\[3mm]
{\it Theory Division, CERN,} \\[1mm]
{\it CH-1211 Geneva 23, Switzerland}\\[1mm]
{ E-mail: schulerg@cernvm.cern.ch}\\[2ex]
{\large and} \\[2ex]
{\Large Torbj\"orn Sj\"ostrand} \\[3mm]
{\it Department of Theoretical Physics,}\\[1mm]
{\it University of Lund, Lund, Sweden}\\[1mm]
{ E-mail: torbjorn@thep.lu.se}
\end{center}
 
\vspace{\fill}
 
\begin{center}
{\bf Abstract}\\[2ex]
\begin{minipage}{\abstwidth}
We propose a generic ansatz for the extension of parton
distributions of the real photon to those of the virtual
photon. Alternatives and approximations are studied that allow
closed-form parametrizations.
\end{minipage}
\end{center}
 
\vspace{\fill}
\noindent
\rule{60mm}{0.4mm}
 
\vspace{1mm} \noindent
${}^a$ Heisenberg Fellow.
 
\vspace{10mm}\noindent
CERN--TH/96--04
 
\clearpage
\pagestyle{plain}
\setcounter{page}{1}
 
\section{Introduction}
 
It is well known that the real photon has a partonic substructure,
induced by virtual fluctuations into $\q\qbar$ pairs. These
fluctuations are in part non-perturbative, and so cannot be
    calculated from first principles (at least not without major advances
    in lattice-gauge theory). Only if the parton distributions
are specified by hand at some sufficiently large input scale $Q_0$
can the continued evolution with $Q^2$ be described perturbatively.
Here $Q^2$ is the scale of the ``probing'' hard process.
    Such a partonic substructure also exists for a spacelike
    photon, $P^2 = - p_{\gamma}^2 > 0$.
Only if $P^2$ is in the deeply inelastic scattering region can the
substructure be neglected; the effects here die away like a
    higher-twist contribution. For the experimentally accessible and
    theoretically challenging region $\Lambda_{\mrm QCD}^2
    \lessim P^2 \lessim 2\,$GeV$^2$ evolution equations for the 
    parton distributions and their boundary conditions cannot be 
     derived from perturbative QCD. In this letter
    we propose a theoretical ansatz 
    to prescribe the modification of the $Q^2$-evolution equations
    of the parton distributions with changing $P^2$ and the input at
    $Q_0$ as a function of $P^2$.
 
While the parton distribution functions (pdf's) of the real photon have
been studied in some detail, both experimentally and theoretically
(for a recent survey see e.g. \cite{LEP2report}), much less is known
about the virtual photon. The only published data are by the PLUTO
Collaboration \cite{PLUTO}. However, recently the ZEUS Collaboration
presented new data from HERA \cite{ZEUS}. The observed $x_{\gamma}$
distribution has been constructed for events with two jets above
    4 GeV transverse momentum. As $P^2$ is increased, this
    distribution is gradually
suppressed at small $x_{\gamma}$, in agreement with theoretical
expectations. These first results will be followed by more with
increasing precision. Also LEP~2 should contribute in the future
\cite{LEP2report}.
 
In a previous publication we presented several parametrizations for
the parton distributions of the real photon \cite{SaS}, with a
proposed extension to virtual photons. This is one of the very few
    studies that give explicit predictions. Drees and Godbole \cite{DG},
    following an analysis of Borzumati and Schuler \cite{Borzu},
have proposed a method based on simple multiplicative factors
    relative to the parton distributions of the real photons.
    A similar recipe  has been used by Aurenche and collaborators
    \cite{AFFGKS}. This may
be useful for estimating the effects of mildly virtual photons in
a sample of almost real photons, but does not appear well suited for
    QCD tests of the virtual-photon distributions. A detailed
study is performed by Gl\"uck, Reya and Stratmann \cite{GRS},
but a main disadvantage here is that there exists no explicit
parametrizations of the resulting distributions. The strategy
adopted is also only one possibility.
 
In this letter we extend our previous study to a few alternative
approaches for the virtual photon, allowing a study of the limits
of our current understanding. Numerical approximations are introduced, 
which permits the parametrizations of real-photon pdf's to be easily
extended to virtual-photon dittos. Section 2 contains a brief
summary on the real photon, the virtual one is covered
in section 3 and some comparisons are shown in section 4.
 
\section{The Real Photon}
 
The parton distributions of the real photon obey a set of
inhomogeneous evolution equations:
\begin{equation}
\frac{\partial f_a^{\gamma}(x,Q^2)}{\partial \ln Q^2} =
\int_x^1 \frac{\d y}{y} \, f_b^{\gamma}(y,Q^2)\,
\frac{\alphas}{2\pi} \, P_{a/b} \left( \frac{x}{y} \right)
+ 3 \, e_a^2 \, \frac{\alphaem}{2\pi} \left( x^2 + (1-x)^2
\right) ~,
\label{gamevol}
\end{equation}
to leading order. The first term is the one present in standard
evolution equations, e.g.\ for the proton. The second term, the
so-called anomalous one, comes from branchings $\gamma \to \q\qbar$,
and is unique for the photon evolution equations.
 
The solution can be written as the sum of two terms,
\begin{equation}
f_a^{\gamma}(x,Q^2) = f_a^{\gamma,\mrm{NP}}(x,Q^2;Q_0^2)
+ f_a^{\gamma,\mrm{PT}}(x,Q^2;Q_0^2) ~.
\end{equation}
The former term is a solution to the homogeneous evolution
(i.e. without the second term in eq.~(\ref{gamevol}))
with a non-perturbative input at $Q=Q_0$, and the latter is a
solution to the full inhomogeneous equation with boundary condition
$f_a^{\gamma,\mrm{PT}}(x,Q_0^2;Q_0^2) \equiv 0$. One possible
physics interpretation is to let $f_a^{\gamma,\mrm{NP}}$ correspond
to $\gamma \leftrightarrow V$ fluctuations, where
$V = \rho^0, \omega, \phi, \mrm{J}/\psi, \ldots$, is a set of vector
mesons, and let $f_a^{\gamma,\mrm{PT}}$ correspond to perturbative
(``anomalous'') $\gamma \leftrightarrow \q\qbar$ fluctuations,
$\q =$ u, d, s, c and b. The discrete spectrum of vector mesons can
be combined with the continuous (in virtuality $k^2$) spectrum of
$\q\qbar$ fluctuations, to give
\begin{equation}
f_a^{\gamma}(x,Q^2) =
\sum_V \frac{4\pi\alphaem}{f_V^2} f_a^{\gamma,V}(x,Q^2; Q_0^2)
+ \frac{\alphaem}{2\pi} \, \sum_{\q} 2 e_{\q}^2 \,
\int_{Q_0^2}^{Q^2} \frac{{\d} k^2}{k^2} \,
f_a^{\gamma,\q\qbar}(x,Q^2;k^2) ~,
\label{decomp}
\end{equation}
where each component $f^{\gamma,V}$ and $f^{\gamma,\q\qbar}$ obeys a
unit momentum sum rule. Although each component formally depends on
two scales, $Q_0^2/k^2$ and $Q^2$, it is the combination
\begin{equation}
s = \int_{Q_0^2/k^2}^{Q^2} \frac{\alphas(r^2)}{2\pi} \,
\frac{{\d} r^2}{r^2}
\label{srange}
\end{equation}
that sets the length of the evolution range and thus gives the
full scale dependence.
 
Beyond this fairly general ansatz, a number of choices has to be
made. The approach adopted in the SaS sets \cite{SaS} is described
in the following.
 
What is $Q_0$? A low scale, $Q_0 \approx 0.6$~GeV, is favoured if
the $V$ states above are to be associated with the lowest-lying
resonances only. Then one expects $Q_0 \sim m_{\rho}/2$--$m_{\rho}$.
Furthermore, with this $Q_0$ one obtains a reasonable description of
the total $\gamma\p$ cross section, and continuity, e.g.\ in the
primordial $k_{\perp}$ spectrum \cite{gammap}.
Against this choice there are worries that perturbation theory may not
be valid at such low $Q$, or at least that higher-twist terms appear
in addition to the standard ones. Alternatively one could
therefore pick a larger value, $Q_0 \approx 2$~GeV, where these
worries are absent. One then needs to include also higher resonances
in the vector-meson sector, which adds some arbitrariness, and one
can no longer compare with low-$Q$ data. We have chosen to prepare
sets for both these (extreme) alternatives.
 
How is the direct contribution to be handled? 
Unlike the p, the $\gamma$ has a
direct component where the photon acts as an unresolved probe.
In the definition of $F_2^{\gamma}$ this adds a component
$C^{\gamma}$, symbolically
\begin{equation}
F_2^{\gamma}(x,Q^2) = \sum_{\q} e_{\q}^2 \left[ f_{\q}^{\gamma} +
f_{\qbar}^{\gamma} \right] \otimes C_{\q} +
f_{\g}^{\gamma} \otimes C_{\g} + C^{\gamma} ~.
\end{equation}
Since $C^{\gamma} \equiv 0$ in leading order, and since we stay with
leading-order fits, it is permissible to neglect this complication.
Numerically, however, it makes a non-negligible difference. We
therefore make two kinds of fits, one DIS type with $C^{\gamma} = 0$
and one $\overline{\mrm{MS}}$ type including the universal part
of $C^{\gamma}$\cite{AFG}.
 
How are heavy flavours, i.e.\ mainly charm, to be dealt with? 
When jet production
is studied for real incoming photons, the standard evolution approach
is reasonable, but with a lower cut-off $Q_0 \approx m_{\c}$ for
$\gamma \to \c\cbar$. Moving to deep inelastic scattering,
$\e\gamma \to \e X$, there is an extra kinematical constraint:
$W^2 = Q^2 (1-x)/x > 4 m_{\c}^2$. It is here better to use the
``Bethe-Heitler'' cross section for $\gamma^* \gamma^* \to \c\cbar$.
Therefore two kinds of output are provided. The parton distributions
are calculated as the sum of a vector-meson part and an anomalous part
including all five flavours, while $F_2^{\gamma}$ is calculated
separately from the sum of the same vector-meson part, an anomalous
part and possibly a $C^{\gamma}$ part now only covering the three
light flavours, and a Bethe-Heitler part for c and b.
 
Should $\rho^0$ and $\omega$ be added coherently or incoherently?
In a coherent mixture, $\u\ubar : \d\dbar = 4 : 1$ at $Q_0$, while
the incoherent mixture gives $1 : 1$. We argue for
coherence at the short distances probed by parton distributions.
This contrasts with long-distance processes, such as
$\p\gamma \to \p V$.
 
    What is $\Lambda_{\mrm{QCD}}$? The $F_2^\gamma$ data are not
    good enough to allow
a precise determination. Therefore we use a fixed value
$\Lambda^{(4)} = 200$~MeV, in agreement with conventional
results for proton distributions.
 
In total, four distributions are presented \cite{SaS}, based on fits
to available data:
\begin{Itemize}
\item SaS 1D, with $Q_0 = 0.6$~GeV and in the DIS scheme.
\item SaS 1M, with $Q_0 = 0.6$~GeV and in the $\overline{\mrm{MS}}$
scheme.
\item SaS 2D, with $Q_0 = 2$~GeV and in the DIS scheme.
\item SaS 2M, with $Q_0 = 2$~GeV and in the $\overline{\mrm{MS}}$
scheme.
\end{Itemize}
The VMD distributions and the integral of the
anomalous distributions are parametrized separately and added
to give the full result; this is of importance for the following.
 
\section{The Virtual Photon}
 
    The evolution equations (in $Q^2$) of the pdf's of the virtual
    photon (and its solutions)
    can be exactly calculated in perturbative QCD for a
    restricted $P^2$ range, namely \cite{Walsh}
    \begin{equation}
    Q_0^2 \ll P^2 \ll Q^2 \ . \label{Prange}
    \end{equation}
    Experimentally accessible, and theoretically challenging is,
    however, the low-$P^2$ range $\Lambda_{\mrm{QCD}}^2 \lessim
    P^2 \lessim Q_0^2$ where evolution equations cannot be derived 
    from perturbative QCD. We propose pdf's that are valid for all $0
    \leq P^2 \leq Q^2$. These are arrived at as follows. We start from
    the observation that the moments of the pdf's are analytic in
    the $P^2$ plane. A natural way to make use of this property is
    \cite{Bj} to express them in terms of a dispersion-integral in
    the (time-like) mass square $k^2$ of the $\q\qbar$ fluctuations.
    This links perturbative and non-perturbative contributions and
    allows the smooth limit $P^2 \rightarrow 0$. The model-dependence
    enters when specifying the necessary weight functions.
    We choose these in such a way that
    the resulting expressions possess the correct, known behaviours
    for both $P^2 \rightarrow 0$ and the range (\ref{Prange}).
    The result is
\begin{equation}
f_a^{\gamma^\star}(x,Q^2,P^2) = \int_0^{Q^2} \frac{{\d} k^2}{k^2}\,
\left( \frac{k^2}{k^2 + P^2} \right)^2 \, \frac{\alphaem}{2\pi} \,
\sum_{\q} 2 e_{\q}^2 \, f_a^{\gamma,\q\qbar}(x,Q^2;k^2) ~.
\label{virtualGVD}
\end{equation}
A definite behaviour for $P^2 \rightarrow Q^2$ has not yet been imposed. 

    The non-appearance of a $1/P^2$ contribution in (\ref{virtualGVD})
    can be argued in two ways. First this is what one expects when
    applying generalized vector-meson dominance to a continuous (in $k^2$)
    spectrum of $\q\qbar$ fluctuations. Second, at large $P^2$ an
    operator-product expansion of the pdf's in powers of $1/P$ holds, 
    but the first non-vanishing higher-twist contribution comes
    from the dimension-$4$ gluon condensate \cite{Gorskii}.
 
    Associating the low-$k^2$ part of relation (\ref{virtualGVD})
    with the discrete
set of vector mesons gives a generalization of
eq.~(\ref{decomp}) to
\begin{eqnarray}
f_a^{\gamma^\star}(x,Q^2,P^2)
& = & \sum_V \frac{4\pi\alphaem}{f_V^2} \left(
\frac{m_V^2}{m_V^2 + P^2} \right)^2 \,
f_a^{\gamma,V}(x,Q^2;\tilde{Q}_0^2)
\nonumber \\
& + & \frac{\alphaem}{2\pi} \, \sum_{\q} 2 e_{\q}^2 \,
\int_{Q_0^2}^{Q^2} \frac{{\d} k^2}{k^2} \, \left(
\frac{k^2}{k^2 + P^2} \right)^2 \, f_a^{\gamma,\q\qbar}(x,Q^2;k^2)
~.
\label{decompvirt}
\end{eqnarray}
In addition to the introduction of the dipole form factors,
note that the lower input scale for the VMD states is here shifted
from $Q_0^2$ to some $\tilde{Q}_0^2 \geq Q_0^2$. This is based on
a study of the evolution equation \cite{Borzu} that shows that
the evolution effectively starts ``later'' in $Q^2$ for a virtual
photon. $\tilde{Q}_0$ can be associated with the $P_0$, $P_0'$,
$P_{\mrm{eff}}$ or $P_{\mrm{int}}$ scales to be introduced below.
 
Equation~(\ref{decompvirt}) is one possible answer, which we will
use as a reference in the following. It depends on both $Q^2$ and
$P^2$ in a non-trivial way, however, so that results are only obtained
by a time-consuming numerical integration rather than as a simple
parametrization.
 
In order to obtain a tractable answer, one may note that the factor
$(k^2/(k^2+P^2))^2$ provides an effective cut-off at $k \approx P$,
so one possible substitution for the anomalous component is
\begin{equation}
\int_{Q_0^2}^{Q^2} \frac{{\d}k^2}{k^2} \, \left(
\frac{k^2}{k^2 + P^2} \right)^2 \, \Big[ \cdots \Big]
     \longmapsto     \int_{P_0^2}^{Q^2} \frac{{\d}k^2}{k^2}
\Big[ \cdots \Big] ~,
\label{SaSapprox}
\end{equation}
with $P_0^2 = \max(Q_0^2,P^2)$. All the $Q^2$ and $P^2$ dependence
is now appearing in a combination like $s$ in eq.~(\ref{srange}),
i.e. parametrizations of $f_a^{\gamma^\star}(x, Q^2, P^2)$ are
readily available by simple modifications of those for $P^2=0$.
This gives the approach we adopted in ref.~\cite{SaS}.
 
Some objections can be raised against this substitution. 
The choice of $P_0$ means that the anomalous component is independent
of $P$ for $P < Q_0$, and that there is a discontinuous change in
behaviour at $P=Q_0$. Other simple expressions could have been adopted
    to solve the problem, such as\footnote{
    In the following
    the $P_0'$ prescription implies also the replacement $Q^2
     \longmapsto Q^2 + P^2\ Q_0^2/Q^2$ in (\ref{decompvirt}) so as to 
    ensure sensible behaviours for $Q^2 \rightarrow Q_0^2$ and 
    $P^2 \rightarrow Q^2$.}
    $P_0'^2 = Q_0^2 + P^2$, but this only
illustrates the arbitrariness of the choice.
 
One guiding principle could be the preservation of the 
momentum sum.
We recall that the components $f^{\gamma,V}$ and
$f^{\gamma,\q\qbar}$ integrate to unit momentum, i.e.
\begin{equation}
\sum_a \int_0^1 {\d}x \, f_a^{\gamma^\star}(x,Q^2,P^2)
 =  \sum_V \frac{4\pi\alphaem}{f_V^2} \left(
\frac{m_V^2}{m_V^2 + P^2} \right)^2 \, +
 \frac{\alphaem}{2\pi} \, \sum_{\q} 2 e_{\q}^2 \,
\int_{Q_0^2}^{Q^2} \frac{{\d} k^2}{k^2} \, \left(
\frac{k^2}{k^2 + P^2} \right)^2
\end{equation}
is a measure of the probability for a photon to be in a
$V$ or $\q\qbar$ state. Momentum sum preservation would thus
suggest the introduction of a scale $P_{\mrm{eff}}^2$
according to
\begin{equation}
\int_{Q_0^2}^{Q^2} \frac{{\d}k^2}{k^2} \, \left(
\frac{k^2}{k^2 + P^2} \right)^2 \equiv
\int_{P_{\mrm{eff}}^2}^{Q^2} \frac{{\d}k^2}{k^2}
~,
\end{equation}
which gives
\begin{equation}
P_{\mrm{eff}}^2 = Q^2 \frac{Q_0^2 + P^2}{Q^2 + P^2}
\exp \left\{ \frac{P^2 (Q^2 -Q_0^2)}{(Q^2 + P^2)(Q_0^2 + P^2)}
\right\} ~.
\end{equation}
For $Q^2 \gg P^2$ this simplifies to
$P_{\mrm{eff}}^2 = (Q_0^2 + P^2) \exp(P^2/(Q_0^2 + P^2))$.
A simple recipe is then to use $P_{\mrm{eff}}$ as a lower cut-off
    for the anomalous components (and also as the 
    expression for $\tilde{Q}_0$
    in $f_a^{\gamma,V}$ in (\ref{decompvirt})).
 
While $P_{\mrm{eff}}$ gives the same normalization of parton
distributions as does eq.~(\ref{decompvirt}), it does not give the
same average evolution range and therefore not the same $x$ shape.
That is, eq.~(\ref{decompvirt}) receives contributions from
components $f^{\gamma,\q\qbar}(x,Q^2,k^2)$ with $k$ down to
$Q_0$, and thus corresponds to a larger average evolution range
$s$ than a procedure with a sharp cut-off $k > P_{\mrm{eff}}$.
In order to reproduce the $x$ shape better, an intermediate
$P_{\mrm{int}}$ in the range $Q_0 < P_{\mrm{int}} < P_{\mrm{eff}}$
is to be preferred. We found no simple formula that defines an optimal
$P_{\mrm{int}}$, so will use $P_{\mrm{int}}^2 = Q_0 P_{\mrm{eff}}$
as a pragmatic choice. The momentum sum can be preserved by a
simple prefactor, i.e. in total the anomalous component is changed
according to
\begin{equation}
\int_{Q_0^2}^{Q^2} \frac{{\d}k^2}{k^2} \, \left(
\frac{k^2}{k^2 + P^2} \right)^2 \, \Big[ \cdots \Big]
     \longmapsto
\frac{\ln(Q^2/P_{\mrm{eff}}^2)}{\ln(Q^2/P_{\mrm{int}}^2)}
\int_{P_{\mrm{int}}^2}^{Q^2} \frac{{\d}k^2}{k^2}
\Big[ \cdots \Big] ~.
\label{Pintpre}
\end{equation}
 
One should note that the above approaches correspond to different
    evolution equations. The pdf's of the virtual photon given by
    our ansatz~(\ref{decompvirt}) obey evolution equations different
    from those of the real
photon, eq.~(\ref{gamevol}): the homogeneous term is the same but
the inhomogeneous one is multiplied by a factor $(Q^2/(Q^2+P^2))^2$.
That is, the branchings $\gamma^\star \to \q\qbar$ are suppressed
    relative to those of a real photon, in accordance with the
    relation (\ref{virtualGVD}). Approximately
    this also holds for the $P_{\mrm{eff}}$ and
    $P_{\mrm{int}}$ prescriptions. The introduction of $P_0$ or $P_0'$
    according to eq.~(\ref{SaSapprox}) removes the $(Q^2/(Q^2+P^2))^2$
    factor, i.e. restores evolution to be fully according to
    eq.~(\ref{gamevol}). Hence differences between the pdf's of the
    real photon and those of the virtual photon in the $P_0$ and $P_0'$
    schemes (and also the difference between the latter two) arise
    solely from different input distributions.
    Evolution equations for parton distributions
    are rigorously defined only in the range (\ref{Prange}), where
    $(Q^2/(Q^2+P^2))^2 \approx 1$. Hence, differences 
    of the kind $Q^2/(Q^2+P^2)$ are formally legitimate,
    but since the evolution normally is started from a $Q_0$ of the
    same order as $P$, they are non-negligible numerically. 

    So far we have not imposed any constraint on the $P^2 \rightarrow
    Q^2$ behaviours of the pdf's of the virtual photon. For $P^2 
    \approx Q^2$ power-like terms $\propto (P^2/Q^2)^p$ are more
    important than the logarithmic ones ($\propto \ln Q^2/P^2$). 
    Then calculations based on fixed-order perturbation
     theory, where
     the full $P^2$ dependence is kept to the order considered,
    are more appropriate than ones invoking pdf's that sum leading
    logarithms. Indeed, for $P^2 \rightarrow Q^2$ resolved-photon
    contributions originating from the quark (gluon) content
    of the virtual photon become part of the $O(\alpha_s)$
    ($O(\alpha_s^2)$) corrections\footnote{That is, the gluon 
    distribution vanishes faster than the quark distributions 
    for $P^2 \rightarrow Q^2$ \cite{Borzu}.}
    to the leading (in $\alpha_s$) direct-photon contributions. 
    A sensible scheme is therefore arrived at by demanding the
    pdf's of the virtual photon to approach the respective 
    parton-model expressions, which vanish like $ \ln Q^2/P^2$
    for the quark distributions and faster for the gluon distribution. 
    Such a behaviour is already respected by the $P_0$ and $P_0'$ schemes. 
     To ensure the same limiting behaviour also for the other two schemes
    we modify the $P_{\mrm{eff}}$ and $P_{\mrm{int}}$
    schemes as follows (recall $P_0^2 = \max(Q_0^2,P^2)$):
\begin{eqnarray}
 P^2_{\mrm{eff}} & \longmapsto &
  \left( 1 - P^2/Q^2 \right) P^2_{\mrm{eff}}
      + \left( P^2 / Q^2 \right) P^2_0 ~;
\nonumber\\
 P^2_{\mrm{int}} & \longmapsto & 
  \left( 1 - P^2/Q^2 \right) P^2_{\mrm{int}}
      + \left( P^2 / Q^2 \right) P^2_{\mrm{eff}}
    \quad \mrm{in}\; \ln(Q^2/P^2_{\mrm{int}}) \;\mrm{in}\;
  \mrm{eq.}~(\ref{Pintpre}) ~;
\nonumber\\
 P^2_{\mrm{int}} & \longmapsto & 
  \left( 1 - P^2/Q^2 \right) P^2_{\mrm{int}}
     + \left( P^2 / Q^2 \right) P^2_0
    \quad \mrm{in}\;\mrm{the}\;k^2\ \mrm{integration}\;\mrm{in}\;
  \mrm{eq.}~(\ref{Pintpre})
\ .
\end{eqnarray} 
    Eventually one will aime at a complete description of 
    photon-induced reactions where resolved-photon contributions 
    (i.e.\ those involving the pdf's of the virtual photon) 
    are matched with direct-photon ones by subtracting from the 
    latter those terms that are already included through the
    pdf's of the photon.

\section{Comparisons}
 
The different approaches studied above may be seen as alternatives,
indicating a spread of uncertainty caused by our limited
understanding. The $P_{\mrm{eff}}$ and $P_{\mrm{int}}$ ones are
attempts to obtain simple numerical approximations to
eq.~(\ref{decompvirt}), while the $P_0$ and $P_0'$ ones are
more loosely related, cf. the comment above on evolution
equations.
 
Further uncertainties come from the assumed pdf's of the
real photon and from the conventional scale ambiguity problems.
(Is $Q$ set by the $p_{\perp}$ of a jet or by some multiple thereof?)
While very significant in their own right, they are not studied
here.
 
The above prescriptions can be applied to any set of parton
distributions, provided that the VMD and anomalous parts have
been parametrized separately. For illustrative purposes, we use
SaS~1D throughout in the following.
The results are similar for set 2D although differences
are reduced in magnitude due to the larger $Q_0$ value.
 
Fig.~1 compares the u quark distribution at $P^2 =0$, 0.25 and
1~GeV$^2$, for $Q^2 =10$~GeV$^2$. All approaches have in common that
the small-$x$ part of the spectrum is reduced more than the
high-$x$ one, proportionally speaking, reflecting the larger
suppression of components with longer evolution range $s$.
The $P_0$ and $P_0'$ approaches are considerably above the
integral of eq.~(\ref{decompvirt}), reflecting the difference
in momentum sum. Among the alternatives with the same momentum sum,
the shorter average evolution range for $P_{\mrm{eff}}$ is reflected
in more quarks at large $x$ (and less gluons), while
$P_{\mrm{int}}$ gives a good approximation to eq.~(\ref{decompvirt}).
(In principle the integral could be performed with either scale choice
for $\tilde{Q}_0$ in the VMD evolution range, but differences are
minor and so we choose to use $\tilde{Q}_0=P_0'$ as a simple
alternative.)
 
ZEUS studies the QCD jet rate for $p_{\perp} > 4$~GeV, i.e. for
$\langle Q^2 \rangle \approx 20$~GeV$^2$, in the range
$0.1 \lessim x < 0.75$ \cite{ZEUS}. Events with $x > 0.75$ 
are assumed to be dominated by the direct
process, which is less sensitive to the $P^2$ variation and is
therefore used to fix the normalization. The ZEUS results cannot be
directly compared with pdf parametrizations, since effects of event
migration and experimental cuts are non-trivial. However, to give some
first impression, we study the quantity
\begin{equation}
\mathcal{I}(P^2; Q^2=20~\mrm{GeV}^2) = \int_{0.1}^{0.75} {\d}x \left\{
\sum_{\q} \left[ xq(x,Q^2;P^2) + x\overline{q}(x,Q^2;P^2) \right]
+ \frac{9}{4} xg(x,Q^2;P^2) \right\} ~,
\label{Idef}
\end{equation}
which is related to the QCD jet rate. The colour factor $9/4$ is the
standard enhancement of gluon interactions relative to quark ones.
The variation of $\mathcal{I}(P^2)/\mathcal{I}(0)$ is shown in
Fig.~2. (It can be discussed whether one should have omitted the $x$
factor in eq.~(\ref{Idef}); one would then have obtained a slightly
steeper fall-off.)
The integral in eq.~(\ref{decompvirt}) and its  $P_{\mrm{eff}}$
and  $P_{\mrm{int}}$ approximations give a drop by a factor of about
2 between $P^2 = 0$ and $P^2 =0.5$~GeV$^2$, in rough agreement with
the ZEUS data, while the $P_0$ and $P_0'$ approaches do not drop at
all as much. (Note the discontinuous derivative in the $P_0$ curve
     at $P = Q_0$, caused by the way $P_0$ is defined. There are also
     smaller kinks visible in the $P_0$ and $P_0'$ curves for 
     $P \approx m_{\mrm{c}} = 1.3$~GeV, related to the way the
     charm threshold is modelled.) With more precise
data and acceptance corrections understood, it should therefore be
possible to discriminate among the alternatives.
 
\section{Summary}
 
    In this letter we have studied the extension of the parton
    distributions of the real photon to those of the virtual one.
    Analyticity in $P^2$ allows us to represent the pdf's of the
    virtual photon as a dispersion integral in the mass of the
    $\q\qbar$ fluctuations. We have obtained an explicit solution for
    the pdf's. Under the assumption of a separation of the $\q\qbar$
    fluctuations in a low-mass, discrete sum of vector-meson states
    and a high-mass, continuous spectrum, various constraints on the
    pdf's provide us with a unique result for $P^2 \ll Q^2$. 
    The final expressions for
    pdf's of the virtual photon $f_a^{\gamma^\star}(x,Q^2,P^2)$
    cannot be given a closed form due to their non-trivial dependence
    on the three variables $x$, $Q^2$, and $P^2$. Therefore we
    constructed a parametrization (the $P_{\mrm{int}}$ prescription)
    which
    allows the $f_a^{\gamma^\star}(x,Q^2,P^2)$ to be very well
    approximated by simple modifications to the parton distributions
    of the real photon, i.e. parametrizations are only needed for the
    $x$ and $Q^2$ dependence. This extension of real-photon pdf's to
    those of a virtual photon can be applied to any set of parton
    distributions, provided that the VMD and anomalous parts are
    available separately. It also gives $F_2^{\gamma^\star}(x,Q^2,P^2)$.
 
    In order to allow for a test of the model-dependence of the
    pdf's of the virtual photon we have constructed three
    prescriptions alternative to $P_{\mrm{int}}$, 
    all provided in closed form. The various
    prescriptions correspond to variants of order $P^2/Q^2$
    in the evolution equations and/or boundary conditions.
    The differences are readily visible in the $P^2$ dependence
    of distributions, so  HERA and LEP~2 should offer the opportunity
    to distinguish between alternatives.
 
A program with the SaS parametrizations modified according to the
prescriptions studied in this paper is available on WWW under\\
http://thep.lu.se/tf2/staff/torbjorn/lsasgam2.

\begin{figure}[p]
\begin{center}
\mbox{\epsfig{figure=virtgam1.eps}}
\end{center}
\caption{The u-quark parton distribution
$xu(x,Q^2 = 10 \, \mrm{GeV}^2)/\alphaem$.
Top full line for $P^2 = 0$, below comparison of five alternatives
for $P^2 = 0.25$ and 1~GeV$^2$. Ordered roughly from top to bottom,
dashed is $P_0$, dotted is $P_0'$, dash-dotted is $P_{\mrm{eff}}$,
large dots is $P_{\mrm{int}}$ and full is the integral in
eq.~(\protect\ref{decompvirt}).  }
\end{figure}
 
\begin{figure}[p]
\begin{center}
\mbox{\epsfig{figure=virtgam2.eps}}
\end{center}
\caption{The fall-off of parton distributions with virtuality $P$
     (we have chosen $P$ as $x$ scale rather than $P^2$, so as to
     better show the small-$P$ region),
normalized to the value at $P^2 = 0$,
$\mathcal{I}(P^2)/\mathcal{I}(0)$, for $Q^2 = 20.25$~GeV$^2$.
Here $\mathcal{I}$, defined by eq.~(\protect\ref{Idef}), is the
colour-factor-weighted sum of parton distributions in the range
$0.1 < x < 0.75$. Curves are labelled as in Fig. 1.} 
\end{figure}

\end{document}